# Paleoclimate Implications for Human-Made Climate Change

James E. Hansen and Makiko Sato

NASA Goddard Institute for Space Studies and Columbia University Earth Institute, New York

ABSTRACT

Paleoclimate data help us assess climate sensitivity and potential human-made climate effects. We conclude that Earth in the warmest interglacial periods of the past million years was less than 1°C warmer than in the Holocene. Polar warmth in these interglacials and in the Pliocene does not imply that a substantial cushion remains between today's climate and dangerous warming, but rather that Earth is poised to experience strong amplifying polar feedbacks in response to moderate global warming. Thus goals to limit human-made warming to 2°C are not sufficient – they are prescriptions for disaster. Ice sheet disintegration is nonlinear, spurred by amplifying feedbacks. We suggest that ice sheet mass loss, if warming continues unabated, will be characterized better by a doubling time for mass loss rate than by a linear trend. Satellite gravity data, though too brief to be conclusive, are consistent with a doubling time of 10 years or less, implying the possibility of multi-meter sea level rise this century. Observed accelerating ice sheet mass loss supports our conclusion that Earth's temperature now exceeds the mean Holocene value. Rapid reduction of fossil fuel emissions is required for humanity to succeed in preserving a planet resembling the one on which civilization developed.

## 1. Introduction

Climate change is likely to be the predominant scientific, economic, political and moral issue of the 21$^{st}$ century. The fate of humanity and nature may depend upon early recognition and understanding of human-made effects on Earth's climate (Hansen, 2009).

Tools for assessing the expected climate effects of alternative levels of human-made changes of atmospheric composition include (1) Earth's paleoclimate history, showing how climate responded to past changes of boundary conditions including atmospheric composition, (2) modern observations of climate change, especially global satellite observations, coincident with rapidly changing human-made and natural climate forcings, and (3) climate models and theory, which aid interpretation of observations on all time scales and are useful for projecting future climate under alternative climate forcing scenarios.

This paper emphasizes use of paleoclimate data to help assess the dangerous level of human interference with the atmosphere and climate. We focus on long-term climate trends of the Cenozoic Era and on Milankovitch (1941) glacial-interglacial climate oscillations. The Cenozoic encompasses a wide range of climates, including a planet without large ice sheets, and it allows study of greenhouse gases as both a climate forcing and a feedback. Glacial-interglacial climate swings, because they are slow enough for Earth to be in near energy balance, allow us to determine accurately the 'fast feedback' climate sensitivity to changing boundary conditions.

We first discuss Cenozoic climate change, which places Milankovitch and human-made climate change in perspective. We then use Milankovitch climate oscillations in a framework that accurately defines climate sensitivity to a natural or human-made climate forcing. We summarize how temperature is extracted from ocean cores to clarify the physical significance of this data record, because, we will argue, ocean core temperature data have profound implications about the dangerous level of human-made interference with global climate. Finally we discuss the temporal response of the climate system to the human-made climate forcing.

## 2. Cenozoic Climate Change

The Cenozoic Era, the time since extinction of dinosaurs at the end of the Cretaceous Era, illustrates the huge magnitude of natural climate change. The early Cenozoic was very warm – indeed, polar regions had tropical-like conditions with alligators in Alaska (Markwick, 1998). There were no large ice sheets on the planet, so sea level was about 70 meters higher than today.

Fig. 1 shows estimated global deep ocean temperature in the Cenozoic, the past 65.5 million years. Deep ocean temperature is inferred from a global compilation of oxygen isotopic abundances in ocean sediment cores (Zachos et al., 2001), with temperature extracted from oxygen isotopes via the approximation of Hansen et al. (2008) as discussed below (section 4). (The data for the entire Cenozoic is available at http://www.columbia.edu/~mhs119/TargetCO2) Deep ocean temperature change is similar to global surface temperature change during the Cenozoic, we will argue, until the deep ocean temperature approaches the freezing point of ocean water. Late Pleistocene glacial-interglacial deep ocean temperature changes (Fig. 1c) are only about two-thirds as large as global mean surface temperature changes (section 4).

Earth has been in a long-term cooling trend for the past 50 million years (Fig. 1a). By approximately 34 Mya (million years ago) the planet had become cool enough for a large ice sheet to form on Antarctica. Ice and snow increased the albedo (literally, the 'whiteness') of that continent, an amplifying feedback that contributed to the sharp drop of global temperature at that time. Moderate warming between 30 and 15 Mya was not sufficient to melt all Antarctic ice. The cooling trend resumed about 15 Mya and accelerated as the climate became cold enough for ice sheets to form in the Northern Hemisphere and provide their amplifying feedback.

The Cenozoic climate changes summarized in Fig. 1 contain insights and quantitative information relevant to assessment of human-made climate effects. Carbon dioxide ($CO_2$) plays a central role in both the long-term climate trends and the Milankovitch oscillations (Fig. 1b) that were magnified as the planet became colder and the ice sheets larger. Cenozoic climate change is discussed by Zachos et al. (2001), IPCC (2007), Hansen et al. (2008), and many others. We focus here on implications about the role of $CO_2$ in climate change and climate sensitivity.

$CO_2$ is the principal forcing that caused the slow Cenozoic climate trends. The total amount of $CO_2$ in surface carbon reservoirs (atmosphere, ocean, soil, biosphere) changes over millions of years due to imbalance of the volcanic source and weathering sink, and changes of the amount of carbon buried in organic matter. $CO_2$ is also a principal factor in the short-term climate oscillations that are so apparent in parts (b) and (c) of Fig. 1. However, in these glacial-interglacial oscillations atmospheric $CO_2$ operates as a feedback: total $CO_2$ in the surface reservoirs changes little on these shorter time scales, but the distribution of $CO_2$ among the surface reservoirs changes as climate changes. As the ocean warms, for example, it releases $CO_2$ to the atmosphere, providing an amplifying climate feedback that causes further warming.

The fact that $CO_2$ is the dominant cause of long-term Cenozoic climate trends is obvious Earth's energy budget. Redistribution of energy in the climate system via changes of atmosphere or ocean dynamics cannot cause such huge climate change. Instead a substantial global climate forcing is required. The climate forcing must be due to a change of energy coming into the planet or changes within the atmosphere or on the surface that alter the planet's energy budget.

Solar luminosity is increasing on long time scales, as our sun is at an early stage of solar evolution, "burning" hydrogen, forming helium by nuclear fusion, slowly getting brighter. The sun's brightness increased steadily through the Cenozoic, by about 0.4 percent according to solar physics models (Sackmann et al., 1993). Because Earth absorbs about 240 W/m$^2$ of solar energy, the 0.4 percent increase is a forcing of about 1 W/m$^2$. This small linear increase of forcing, by itself, would have caused a modest global warming through the Cenozoic Era.

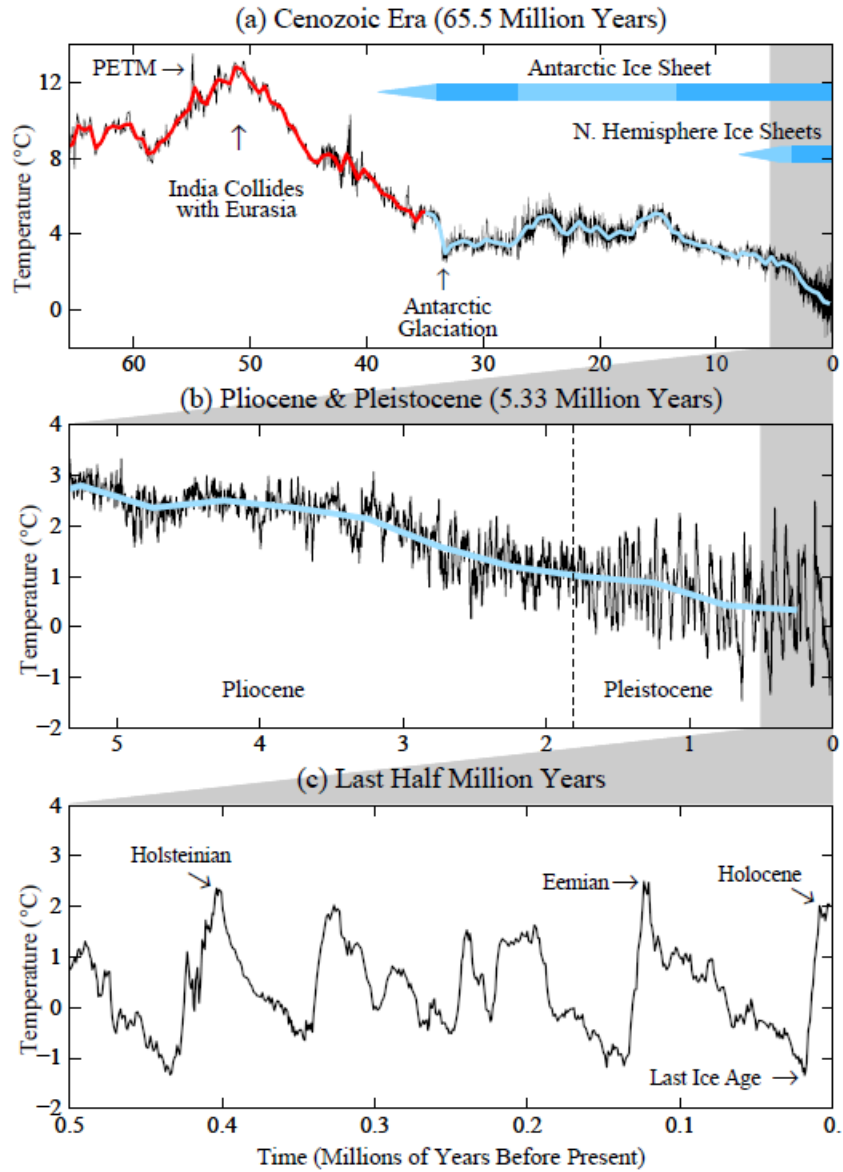

**Fig. 1.** Estimated Cenozoic global deep ocean temperature. Pliocene/Pleistocene is expanded in (b) and the last half million years in (c). High frequency variations (black) are 5-point running means of original data (Zachos et al., 2001); red and blue curves have 500 ky resolution. PETM is the Paleocene Eocene Thermal Maximum. Blue bars indicate ice sheet presence, with dark blue for ice sheets near full size. Holsteinian and Eemian are known in paleoclimate literature as Marine Isotope Stages 11 and 5e.

Continent locations affect Earth's energy balance, as ocean and continent albedos differ. However, most continents were near their present latitudes by the early Cenozoic (Blakey, 2008; Fig. S9 of Hansen et al., 2008). Cloud and atmosphere shielding limit the effect of surface albedo change (Hansen et al., 2005), so this surface climate forcing did not exceed about 1 W/m$^2$.

In contrast, atmospheric $CO_2$ during the Cenozoic changed from about 1000 ppm in the early Cenozoic (Beerling and Royer, 2011) to as small as 170 ppm during recent ice ages (Luthi et al., 2008). The resulting climate forcing, which can be computed accurately for this $CO_2$ range using formulae in Table 1 of Hansen et al. (2000), exceeds 10 W/m$^2$. $CO_2$ was clearly the dominant climate forcing in the Cenozoic.

Global temperature change in the first half of the Cenozoic is consistent with expected effects of plate tectonics (continental drift) on atmospheric $CO_2$. Subduction of ocean crust by an overriding tectonic plate causes crustal melt and metamorphism of the subducted plate and sediments, with release of volatiles including $CO_2$. Carbon amount in surface reservoirs depends on the balance between this outgassing (via volcanoes and seltzer springs) from Earth's crust and burial in the crust, including change in the amount of buried organic matter (Berner, 2004). $CO_2$ outgassing occurs during subduction of oceanic crust and weathering (oxidation) of previously buried organic matter. Burial is via chemical weathering of rocks with deposition of carbonates on the ocean floor and burial of organic matter, some of which eventually may form fossil fuels.

Rates of outgassing and burial of $CO_2$ are each typically $10^{12}$-$10^{13}$ mol C/year (Staudigel et al., 1989; Edmond and Huh, 2003; Berner, 2004). Imbalance between outgassing and burial is limited by negative feedbacks in the geochemical carbon cycle (Berner and Caldeira, 1997), but a net natural imbalance of the order of $10^{12}$ mol C/year can be maintained on long time scales, as continental drift affects the rate of outgassing. Such an imbalance, after distribution among surface reservoirs, is only ~0.0001 ppm/year of atmospheric $CO_2$. That rate is negligible compared to the present human-made atmospheric $CO_2$ increase of ~2 ppm/year, yet in a million years such a consistent crustal imbalance can alter atmospheric $CO_2$ by ~100 ppm.

India was the only land area located far from its current location at the beginning of the Cenozoic. The Indian plate was still south of the Equator, but moving northward at a rate of about 20 cm per year (Kumar et al., 2007), a rapid continental drift rate. The Indian plate moved through the Tethys Ocean, now the Indian Ocean, which had long been the depocenter for carbonate and organic sediments from major world rivers.

The strong global warming trend between 60 and 50 My ago was presumably a consequence of increasing atmospheric $CO_2$, as the Indian plate subducted carbonate-rich ocean crust while traversing the Tethys Ocean (Kent and Muttoni, 2008). The magnitude of the $CO_2$ source continued to increase until India crashed into Asia and began pushing up the Himalaya Mountains and Tibetan Plateau. Emissions from this tectonic source continue even today, but the magnitude of emissions began decreasing after the Indo-Asian collision and as a consequence the planet cooled. The climate variations between 30 and 15 million years ago, when the size of the Antarctic ice sheet fluctuated, may have been due to temporal variations of plate tectonics and outgassing rates (Patriat et al., 2008). Although many mechanisms probably contributed to climate change through the Cenozoic Era, it is clear that $CO_2$ change was the dominant cause of the early warming and the subsequent long-term cooling trend.

Plate tectonics today is producing relatively little subduction of carbonate-rich ocean crust (Edmund and Huh, 2003; Gerlach, 2011), consistent with low Pleistocene levels of $CO_2$ (170-300 ppm) and the cool state of the planet, with ice sheets in the polar regions of both hemispheres. Whether Earth would have cooled further in the absence of humans[1], on time scales of millions of years, is uncertain. But that is an academic question. The rate of human-made change of atmospheric $CO_2$ amount is now much larger than slow geological changes. Humans now determine atmospheric composition, for better or worse, and they are likely to continue to do so, as long as the species survives.

The Cenozoic Era helps us determine the dangerous level of human-made climate change. However, implications of Cenozoic climate change become clearer if we first discuss empirical data on climate sensitivity provided by recent Milankovitch climate oscillations.

---

[1] Paleoanthropological evidence of Homo sapiens in Africa dates to about 200,000 years ago, i.e., over two glacial cycles. Earlier human-like populations, such as Neanderthals and Homo erectus, date back at least 2,000,000 years, but, as is clear from Fig. 1a, even the human-like species were present only during the recent time of ice ages.

## 3. Climate Sensitivity

A climate forcing is an imposed perturbation of Earth's energy balance. Natural forcings include changes of solar irradiance and volcanic aerosols that scatter and absorb solar and terrestrial radiation. Human-made forcings include greenhouse gases (GHGs) and tropospheric aerosols, i.e., aerosols in Earth's lower atmosphere, mostly in the lowest few kilometers.

A forcing, F, is measured in watts per square meter (W/m$^2$) averaged over the planet. For example, if the sun's brightness increases 1 percent the forcing is F ~ 2.4 W/m$^2$, because Earth absorbs about 240 W/m$^2$ of solar energy averaged over the planet's surface. If the $CO_2$ amount in the air is doubled[2], the forcing is F ~ 4 W/m$^2$. This $CO_2$ forcing is obtained by calculating its effect on the planetary energy balance with all other atmospheric and surface properties fixed. The $CO_2$ opacity as a function of wavelength is known from basic quantum physics and verified by laboratory measurements to an accuracy of a few percent. No climate model is needed to calculate the forcing. It requires only summing over the planet the change of heat radiation to space, which depends on known atmospheric and surface properties.

Climate sensitivity (S) is the equilibrium global surface temperature change (ΔTeq) in response to a specified unit forcing after the planet has come back to energy balance,

$$S = \Delta Teq/F, \quad (1)$$

i.e., climate sensitivity is the eventual (equilibrium) global temperature change per unit forcing.

Climate sensitivity depends upon climate feedbacks, the many physical processes that come into play as climate changes in response to a forcing. Positive (amplifying) feedbacks increase the climate response, while negative (diminishing) feedbacks reduce the response.

Climate feedbacks are the core of the climate problem. Climate feedbacks can be confusing, because, in climate analyses, what is sometimes a climate forcing is other times a climate feedback. As a preface to quantitative evaluation of climate feedbacks and climate sensitivity, we first make a remark about climate models and then briefly summarize Earth's recent climate history to provide specificity to the concept of climate feedbacks.

Climate models, based on physical laws that describe the structure and dynamics of the atmosphere and ocean, as well as processes on land, have been developed to simulate climate. Models help us understand climate sensitivity, because we can change processes in the model one-by-one and study their interactions. But if models were our only tool, climate sensitivity would always have large uncertainty. Models are imperfect and we will never be sure that they include all important processes. Fortunately, Earth's history provides a remarkably rich record of how our planet responded to climate forcings in the past. Paleoclimate records yield, by far, our most accurate assessment of climate sensitivity and climate feedbacks.

Now let us turn to a more general discussion of climate feedbacks, which determine climate sensitivity. Feedbacks do not come into play coincident with a forcing. Instead they occur in response to climate change. It is assumed that, to a useful approximation, feedbacks affecting the global mean response are a function of global temperature change.

'Fast feedbacks' appear almost immediately in response to global temperature change. For example, as Earth becomes warmer the atmosphere holds more water vapor. Water vapor is an amplifying fast feedback, because water vapor is a powerful greenhouse gas. Other fast feedbacks include clouds, natural aerosols, snow cover and sea ice.

---

[2] $CO_2$ climate forcing is approximately logarithmic, because its absorption bands saturate as $CO_2$ amount increases. An equation for climate forcing as a function of $CO_2$ amount is given in Table 1 of Hansen et al. (2000).

'Slow feedbacks' may lag global temperature change by decades, centuries, millennia, or longer time scales. Principal slow feedbacks are surface albedo and long-lived GHGs. It thus turns out that slow feedbacks on millennial time scales are predominately amplifying feedbacks. As a result, the feedbacks cause huge climate oscillations in response to minor perturbations of Earth's orbit that alter the geographical and seasonal distribution of sunlight on Earth.

Surface albedo refers to continental reflectivity. Changes of ice sheet area, continental area, or vegetation cover affect surface albedo and temperature. Hydrologic effects associated with vegetation change also can affect global temperature. Numerical experiments (Hansen et al., 1984) indicate that ice sheet area is the dominant surface feedback in glacial to interglacial climate change, so ice sheet area is a useful proxy for the entire slow surface feedback in Pleistocene climate variations. Surface albedo is an amplifying feedback, because the amount of solar energy absorbed by Earth increases when ice and snow area decreases.

GHGs are also an amplifying feedback on millennial time scales, as warming ocean and soils drive more $CO_2$, $CH_4$ and $N_2O$ into the air. This GHG feedback exists because the atmosphere exchanges carbon and nitrogen with other surface reservoirs (ocean, soil, biosphere).

Negative carbon cycle feedbacks occur, especially on long time scales, via exchange of carbon with the solid earth (Berner, 2004; Archer, 2005). Chemical weathering of rocks, with deposition of carbonates on the ocean floor, slowly removes from surface reservoirs $CO_2$ that is in excess of the amount in equilibrium with natural tectonic (volcanic) $CO_2$ sources. Weathering is thus a diminishing feedback. Unfortunately, the weathering feedback is substantial only on millennial and longer time scales, so it does not alter much the human-made perturbation of atmospheric $CO_2$ on time scales that are of most interest to humanity.

### 3.1. Milankovitch climate oscillations

The glacial-interglacial climate oscillations manifest in Fig.1b and 1c, which grow in amplitude through the Pliocene and Pleistocene, are often referred to as Milankovitch climate oscillations. Milankovitch (1941) suggested that these climate swings occur in association with periodic perturbations of Earth's orbit by other planets (Berger, 1978) that alter the geographical and seasonal distribution of insolation over Earth's surface.

The varying orbital parameters are (1) tilt of Earth's spin axis relative to the orbital plane, (2) eccentricity of Earth's orbit, (3) day of year when Earth is closest to the sun, also describable as precession of the equinoxes (Berger, 1978). These three orbital parameters vary slowly, the dominant time scales being close to 40,000, 100,000 and 20,000 years, respectively.

Hays et al. (1976) confirmed that climate oscillations occur at the frequencies of the periodic orbital perturbations. Wunsch (2003) showed that the dominant orbital frequencies account for only a fraction of total long-term climate variability. That result is not surprising given the small magnitude of the orbital forcing. The orbital forcing, computed as the global-mean annual-mean perturbation of absorbed solar radiation with fixed climate, is less than ±0.25 W/m$^2$ (Fig. S3 of Hansen et al., 2008). Climate variability at other frequencies in the observational data is expected, because orbital changes are more complex than three discrete time scales and because the dating of observed climate variations is imprecise. But it is clear that a large global climate response to the weak orbital forcing does exist (Roe, 2006), demonstrating that climate is very sensitive on millennial time scales and implying that large amplifying feedbacks exist on such time scales. Thus large climate change should also be expected in response to other weak forcings and climate noise (chaos).

A satisfactory quantitative interpretation of how each orbital parameter alters climate has not yet been achieved. Milankovitch argued that the magnitude of summer insolation at high latitudes in the Northern Hemisphere was the key factor determining when glaciation and deglaciation occurred. Huybers (2006) points out that insolation integrated over the summer is affected only by axial tilt. Hansen et al. (2007a) argue that late spring (mid-May) insolation is the key, because early 'flip' of ice sheet albedo to a dark wet condition produces a long summer melt season; they buttress this argument with data for the timing of the last two deglaciations (Termination I 13-14,000 years ago and Termination II about 130,000 years ago).

Fortunately, it is not necessary to have a detailed quantitative theory of the ice ages in order to extract vitally important information. In the following section we show that Milankovitch climate oscillations provide our most accurate assessment of climate sensitivity.

### 3.2. Fast-feedback climate sensitivity

Fast-feedback climate sensitivity can be determined precisely from paleoclimate data for recent glacial-interglacial climate oscillations. This is possible because we can readily find times when Earth was in quasi-equilibrium with its 'boundary forcings'. Boundary forcings are factors that affect the planet's energy balance, such as solar irradiance, continental locations, ice sheet distribution, and atmospheric amount of long-lived GHGs ($CO_2$, $CH_4$ and $N_2O$).

Quasi-equilibrium means Earth is in radiation balance with space within a small fraction of 1 $W/m^2$. For example, the mean planetary energy imbalance was small averaged over several millennia of the Last Glacial Maximum (LGM, which peaked about 20,000 years ago) or averaged over the Holocene (prior to the time of large human-made changes). This assertion is proven by considering the contrary: a sustained imbalance of 1 $W/m^2$ would have melted all ice on Earth or changed ocean temperature a large amount, neither of which occurred.

The altered boundary conditions that maintained the climate change between these two periods had to be changes on Earth's surface and changes of long-lived atmospheric constituents, because the incoming solar energy does not change much in 20,000 years. Changes of long-lived GHGs are known accurately for the past 800,000 years from Antarctic ice core data (Luthi et al., 2008; Loulergue et al., 2008). Climate forcings due to GHG and surface albedo changes between the LGM and Holocene were approximately 3 and 3.5 $W/m^2$, respectively, with largest uncertainty (±1 $W/m^2$) in the surface change (ice sheet area, vegetation distribution, shoreline movement) due to uncertainty in ice sheet sizes (Hansen et al., 1984; Hewitt and Mitchell, 1997).

Global mean temperature change between the LGM and Holocene has been estimated from paleo temperature data and from climate models constrained by paleo data. Shakun and Carlson (2010) obtain a data-based estimate of 4.9°C for the difference between the Altithermal (peak Holocene warmth, prior to the past century) and peak LGM conditions. They suggest that this estimate may be on the low side, mainly because they lack data in some regions where large temperature change is likely, but their record is affected by LGM cooling of 17°C on Greenland. A comprehensive multi-model study of Schneider von Deimling et al. (2006) finds a temperature difference of 5.8 ± 1.4°C between LGM and the Holocene, with this result including the effect of a prescribed LGM aerosol forcing of –1.2 $W/m^2$. The appropriate temperature difference for our purposes is between average Holocene conditions and LGM conditions averaged over several millennia. We take 5 ± 1°C as our best estimate. Although the estimated uncertainty is necessarily partly subjective, we believe it is a generous (large) estimate for 1σ uncertainty.

The empirical fast-feedback climate sensitivity that we infer from the LGM-Holocene comparison is thus 5°C/6.5 $W/m^2$ ~ ¾ ± ¼ °C per $W/m^2$ or 3 ± 1°C for doubled $CO_2$. The fact that ice sheet and GHG boundary conditions are actually slow climate feedbacks is irrelevant for the purpose of evaluating the fast-feedback climate sensitivity.

This empirical climate sensitivity incorporates all fast response feedbacks in the real-world climate system, including changes of water vapor, clouds, aerosols, aerosol effects on clouds, and sea ice. In contrast to climate models, which can only approximate the physical processes and may exclude important processes, the empirical result includes all processes that exist in the real world – and the physics is exact.

If Earth were a blackbody without climate feedbacks the equilibrium response to 4 W/m$^2$ forcing would be about 1.2°C (Hansen et al., 1981, 1984; Lacis et al., 2010), implying that the net effect of all fast feedbacks is to amplify the equilibrium climate response by a factor 2.5. GISS climate models suggest that water vapor and sea ice feedbacks together amplify the sensitivity from 1.2°C to 2-2.5°C. The further amplification to 3°C is the net effect of all other processes, with the most important ones probably being aerosols, clouds, and their interactions.

The empirical sensitivity 3 ± 1°C for doubled $CO_2$ is consistent with the Charney et al. (1979) estimates of 3 ± 1.5°C for doubled $CO_2$ and with the range of model results, 2.1-4.4°C, in the most recent IPCC report (Randall and Wood, 2007). However, the empirical result is more precise, and we can be sure that it includes all real-world processes. Moreover, by examining observed climate change over several Milankovitch oscillations we can further improve the accuracy of the fast-feedback climate sensitivity.

Fig. 2 shows atmospheric $CO_2$ and $CH_4$ and sea level for the past 800,000 years and resulting calculated climate forcings. Sea level implies the total size of the major ice sheets, which thus defines the surface albedo forcing as described by Hansen et al. (2008). Note that calculation of climate forcings due to GHG and ice sheet changes is a radiative calculation; it does not require use of a global climate model. Clouds and other fast-feedback variables are fixed with modern distributions. We do not need to know paleo clouds and aerosols, because the changes of those quantities at earlier climates are in the fast feedback being evaluated.

Multiplying the sum of greenhouse gas and surface albedo forcings by climate sensitivity ¾°C per W/m$^2$ yields the predicted global temperature change (blue curves in Fig. 2d and 2e). Observed temperature change in Fig. 2d is from Dome C in Antarctica (Jouzel et al., 2007). The global deep ocean temperature record in Fig. 2e is from data of Zachos et al. (2001), with temperature extracted from oxygen isotope data as described below and by Hansen et al. (2008).

Observed Antarctic and deep ocean temperature changes have been multiplied by factors (0.5 and 1.5, respectively) to yield observed LGM-Holocene global temperature change of 5°C. Climate sensitivity ¾°C per W/m$^2$ provides a good fit to the entire 800,000 years. An exception is Dome C during the warmest interglacial periods, when warming was greater than calculated. We show in section 4 that peak interglacial warming was probably confined to the ice sheets, so deep ocean temperature change provides a better indication of global temperature change.

The close fit of observed and calculated temperatures for 800,000 years includes multiple tests and thus reduces uncertainty of the implied climate sensitivity. The greatest uncertainty is in the actual global temperature changes. Including our partly subjective estimate of uncertainty, our inferred climate sensitivity is or 3 ± 0.5C for doubled $CO_2$ (3/4 ± 1/8 °C per W/m$^2$).

Regardless of the exact error-bar, this empirically-derived fast-feedback sensitivity has a vitally important characteristic: it incorporates all real-world fast-feedback processes. No climate model can make such a claim.

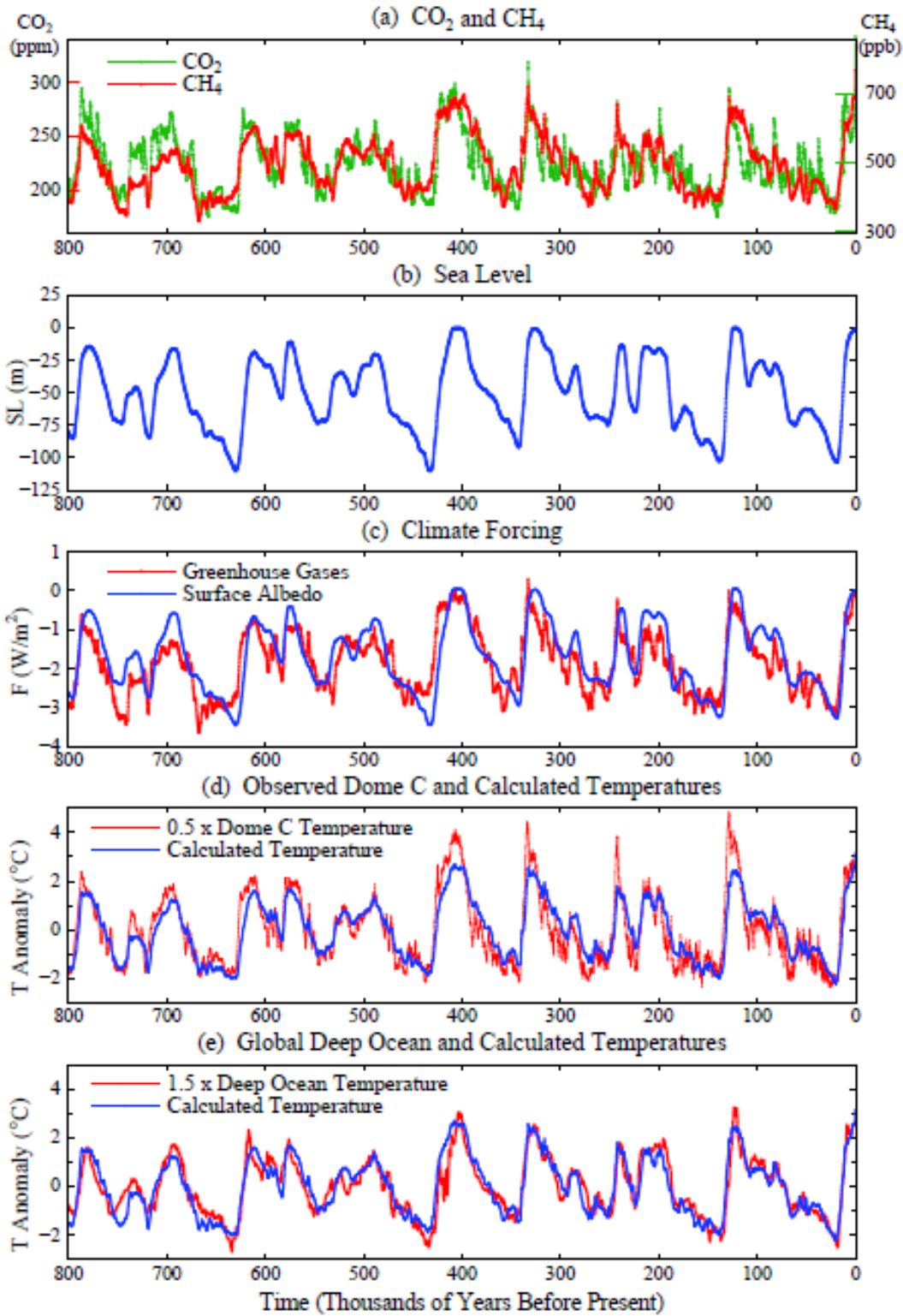

**Fig. 2.** (a) $CO_2$ (Luthi et al., 2008) and $CH_4$ (Loulergue et al., 2008) for past 800,000 years, (b) sea level (Bintanja et al., 2005), (c) resulting climate forcings, (d, e) calculated global temperature anomalies compared with $0.5 \times$ Antarctic Dome C and $1.5 \times$ deep ocean temperatures. Calculations are the product of the forcing and sensitivity ¾°C per W/m². Anomalies are relative to the 800,000 year mean.

## 3.3. Charney climate sensitivity and aerosols

The high precision of the empirical fast-feedback climate sensitivity seems to be at odds with many other climate sensitivity estimates in the scientific literature. Explanation requires background information and clarification of terminology.

Charney et al. (1979), in an early study of climate sensitivity, focused on climate change on the century time scale. Ice sheets were assumed to be fixed and changes of long-lived GHGs were taken as specified climate forcings. In reality, long-lived GHGs are altered by climate change, i.e., there is a GHG feedback effect, but Charney assumed that the feedback change of GHGs would be calculated or estimated separately. This approach, treating ice sheets and long-lived GHGs as fixed boundary conditions or forcings, is an invaluable gedanken experiment and analysis approach, as we have discussed in this paper – even though we know that ice sheets and GHGs will begin to change in response to climate change well before a new fast-feedback climate equilibrium can be achieved.

Charney et al. (1979) used climate models to estimate climate sensitivity. The models included fast feedbacks due to changes of water vapor, clouds and sea ice, but not other fast feedbacks such as changes of aerosols and tropospheric ozone. This landmark study has provided guidance for further studies for decades. But unfortunately the terminology 'Charney sensitivity' has come to be used for multiple definitions of climate sensitivity. Does Charney sensitivity include all fast feedbacks, as we have above, or does it include only the fast feedbacks in the models employed in the Charney study?

Specifically, are glacial-interglacial aerosol changes considered to be a boundary forcing or a fast feedback? In models it is possible, and useful, to turn individual feedbacks on or off – but it is necessary to make clear which feedbacks are included. Similarly, when climate sensitivity is inferred empirically from records of past climate change, it is essential to define which boundary conditions have been defined as climate forcings.

Moreover, the <u>all</u> fast-feedback climate sensitivity has special importance. First, observed climate change necessarily includes all fast feedbacks. Second, it is only the <u>all</u> fast-feedback climate sensitivity that can be derived precisely from paleoclimate records.

Unfortunately, Hansen et al. (1984) chose to estimate climate sensitivity from paleoclimate data by treating the aerosol change between glacial and interglacial conditions as a forcing. There is nothing inherently wrong with asking the question: what is the sensitivity of the remaining processes in the system if we consider ice sheets, GHGs, and aerosols to be specified forcings, even though the ice sheets and GHGs are slow feedbacks and aerosol changes are a fast feedback. The problem is that it is impossible to get an accurate answer to that question. The aerosol forcing depends sensitively on aerosol absorption (the aerosol single scatter albedo) and on the altitude distribution of the aerosols, but, worse, it depends on how the aerosols modify cloud properties. The large uncertainty in the value of the aerosol forcing causes the resulting empirical climate sensitivity to have a large error bar.

Chylek and Lohmann (2008), for example, estimate the aerosol forcing between the last glacial maximum and the Holocene to be 3.3 W/m$^2$, and they thus infer that climate sensitivity for doubled $CO_2$ is 1.8 ± 0.5°C for doubled $CO_2$. With the same approach, but assuming a dust forcing of 1.9 W/m$^2$, Kohler et al. (2010) conclude that climate sensitivity is in the range 1.4–5.2°C for doubled $CO_2$. Both of these studies consider only dust aerosols, so other aerosols are implicitly treated as a climate feedback. Neither study includes aerosols such as black soot, organic particles and dimethyl sulfide (Charlson et al., 1987), whose changes are potentially significant on paleoclimate time scales. Furthermore, neither study includes aerosol indirect forcings, i.e., the effect of aerosols on cloud albedo and cloud cover. IPCC (2007) estimates that the aerosol indirect forcings exceed the direct aerosol forcing, but with a very large uncertainty. Thus interpretation of an empirical climate sensitivity that treats natural aerosol changes as a forcing is complex, and the error bar on the derived sensitivity is necessarily large.

    Also an empirical climate sensitivity that mixes fast and slow processes is less useful for climate analyses. Ice sheet change and natural $CO_2$ change are necessarily slow, while aerosol amount and composition adjust rapidly to climate change. Of course there are aerosol changes on long times scales, for example, some periods are dustier than others. But these aerosol changes are analogous to the cloud changes that occur between climates with or without an ice sheet. Changed surface conditions (e.g., ice sheet area, vegetation cover, land area and continental shelf exposure) cause clouds and aerosols to exhibit changes over long time scales, but the adjustment time of clouds and aerosols to surface conditions is fast.

    Clearly aerosol changes should be included as part of the fast feedback processes in most climate analyses. It makes sense to pull aerosols out of the fast feedbacks only when one is attempting to evaluate the specific contribution of aerosols to the net all-fast-feedback sensitivity. But with such a separation it must be recognized that the error bars will be huge.

    Henceforth, by fast-feedback climate sensitivity, $S_{ff}$, we refer to the <u>all</u> fast-feedback sensitivity. $S_{ff}$ is thus the fast-feedback sensitivity that we estimated from empirical data to be

$$S_{ff} \ = \ 0.75 \pm 0.125 \ °C \ \ per \ W/m^2, \qquad (2)$$

which is equivalent to $3 \pm 0.5°C$ for doubled $CO_2$. High precision is possible for fast-feedback climate sensitivity because GHG amount is known accurately, sea level is known within 20 m, and conversion of sea level change to surface albedo forcing between glacial and interglacial states is not very sensitive to sea level uncertainties (Hansen et al., 2008).

    Climate sensitivity studies that include aerosols as a boundary forcing should use specific appropriate nomenclature. For example, $S_{ff-a}$ can be used to indicate that aerosols are not included in the fast feedbacks. However, it is also necessary to define which aerosols are included as boundary forcings and whether indirect aerosol forcings are included as part of the boundary forcing. Studies evaluating $S_{ff-a}$ can also readily report the implied value for the fast-feedback climate sensitivity, $S_{ff}$. It would be helpful if that information were included for the sake of clarity and comparison with other studies.

    If the terminology 'Charney sensitivity' is to be retained, we suggest that it be reserved for the fast-feedback sensitivity, $S_{ff}$. This all-fast-feedback sensitivity is the logical building block for climate sensitivity on longer time scales as successive slow processes are added.

### 3.4. Slow climate feedbacks

    Fig. 2 shows that glacial-to-interglacial global temperature change is accounted for by changing GHGs and surface albedo. Changes of these boundary forcings affect Earth's temperature by altering the amount of sunlight absorbed by the planet and the amount of heat radiated to space. However, the millennial climate swings were not initiated by GHG and surface albedo changes. Changes of these two boundary forcings were slow climate feedbacks that magnified the climate change. This role is confirmed by the fact that temperature turning points precede the GHG and surface albedo maxima and minima (Mudelsee, 2001). This sequencing is as expected. For example, as the climate warms it is expected that the area of ice and snow will decline, and it is expected that the ocean and continents will release GHGs.

Fig. 3 examines the relation of GHG and surface albedo boundary forcings with global temperature during the past 800,000 years. Each dot is a 1000-year mean temperature anomaly (relative to the most recent 1000 years) plotted against total (GHG + surface albedo) forcing in the upper row, against GHG forcing in the middle row, and against surface albedo forcing in the bottom row. (Surface albedo forcing was computed using the non-linear two-ice-sheet model shown in Fig. S4 of Hansen et al., 2008, but results were indistinguishable for the linear model in that figure.) Temperatures in the left column are from the Dome C Antarctic ice core (Jouzel et al., 2007). Temperatures in the right column are from ocean sediment cores (see section 4).

Dome C temperatures are multiplied by 0.5 and deep ocean temperatures by 1.5 in Fig. 3 so that resulting temperatures approximate global mean temperature. These scale factors were chosen based on the LGM-Holocene global temperature change, as discussed above.

Fig. 3 reveals that the GHG and surface albedo feedbacks increase approximately linearly as a function of global temperature. Moderate nonlinearity of the Dome C temperature, i.e., the more rapid increase of temperature as it approaches the modern value, confirms our contention that deep ocean temperature is a better measure of global temperature change than Antarctic temperature. That conclusion is based on the fact that the temperature changes in Fig. 3 are a result of the fast feedback climate change that is maintained by the changing boundary forcings (GHG amount and ice sheet area). Fast feedback climate sensitivity is nearly linear until Earth approaches either the snowball Earth or runaway greenhouse climate states (Fig. S2 of Hansen et al., 2008). The upturn of Dome C temperatures as a function of boundary forcing is not an indication that Earth is approaching a runaway greenhouse effect. Instead it shows that the Dome C temperature does not continue to be proportional to global mean temperature by a constant factor when Earth is near present day and higher temperatures.

The conclusion that Dome C temperature change cannot be taken today as simply proportional to global temperature change has practical implications. One implication, discussed in section 5, is that a target of 2°C for limiting human-made climate change is too high. We must check the sea level record (Fig. 2b) used to obtain surface albedo forcing, because that sea level curve is based in part on an ice sheet model (Bintanja, et al., 2005). The ice sheet model helps separate contributions of ice volume and deep ocean temperature, which both affect the oxygen isotope record in ocean sediment cores. Our reason for caution is that ice sheet models may be too lethargic, responding more slowly to climate change than real world ice sheets (Hansen, 2005, 2007; Hansen et al., 2007a). We use the Bintanja et al. (2005) sea level data set because it is reasonably consistent with several other sea level data records for the past 400,000 years that do not depend on an ice sheet model (Fig. 2a of Hansen et al. 2007a), and it provides a data set that covers the entire 800,000 years of the Dome C Antarctica record. However, there is one feature in the surface albedo versus temperature scatter plots (Figs. 3e and 3f) that seems unrealistic: the tail at the warmest temperatures, where warming of 1°C produces no change of sea level or surface albedo.

Our check consists of using an independent sea level record based on water residence times in the Red Sea (Siddall et al., 2003). The Sidall et al. data are compared with other sea level records in Fig. 2 of Hansen et al. (2007a) and with GHG and temperature records in Fig. 1 of Hansen et al. (2008). The Siddall et al. (2003) data necessarily cause the scatter-plot (surface albedo versus deep ocean temperature) to become noisier because of inherent imprecision in matching the different time scales of deep ocean temperature and sea level from Red Sea data, but that increased scatter does not obviate the check that we seek.

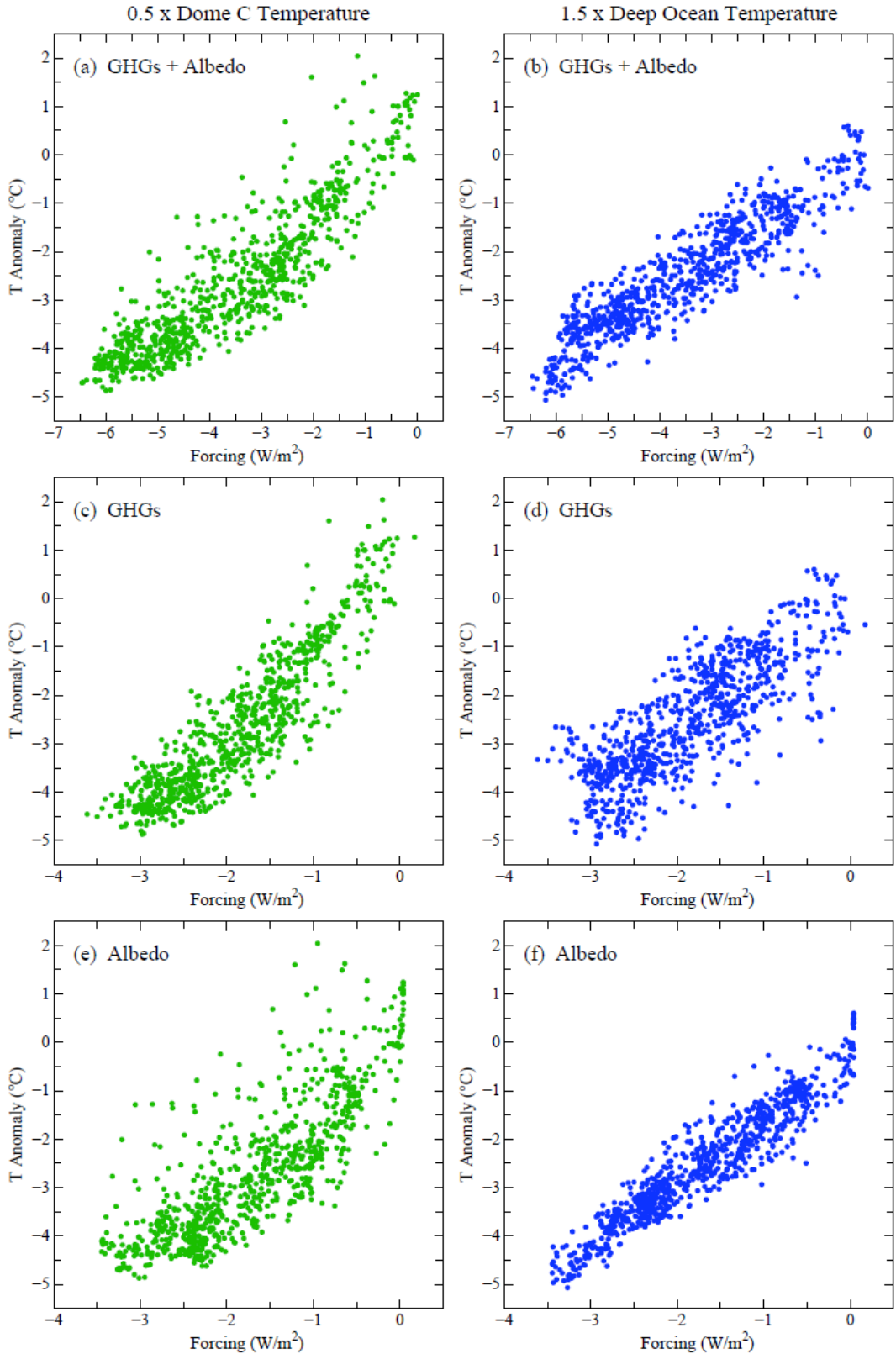

**Fig. 3.** Dome C and deep ocean temperature plotted versus GHG and surface albedo forcings for nominally the same time. Each point is a 1000-year mean from the past 800,000 years (see text).

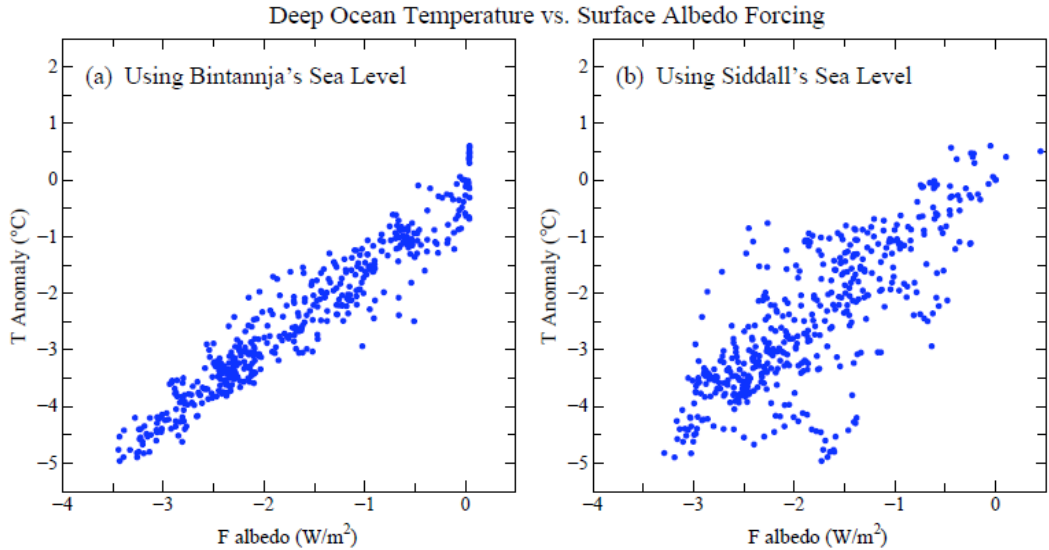

**Fig. 4.** Deep ocean temperature anomalies for the past 470,000 years relative to the past millennium. Each point is the average anomaly over 1000 years plotted against the surface albedo climate forcing calculated from sea level records of Bintanja et al. (2005) for the same 1000 years. Deep ocean anomalies are multiplied by 1.5 to approximate global temperature anomalies.

Fig. 4 confirms the principal characteristic of the Bintaja et al. (2005) sea level data set: a nearly linear relation between deep ocean temperature and sea level. Fig. 4 also confirms our suspicion that the absence of significant sea level response to temperature increase at current temperatures is an artifact, suggesting that the ice sheet model is excessively lethargic. The data not affected by an ice sheet model (Fig. 4b) give no indication of a change in the linear relation of about 20 m equilibrium sea level rise for each 1°C increase of global mean temperature.

### 3.5. Climate sensitivity including slow feedbacks

Climate sensitivity including slow feedbacks is now frequently described as 'Earth system sensitivity' (Lunt et al., 2010; Pagani et al., 2010; Park and Royer, 2011; Royer et al., 2011), but not always with the same definition. There are merits in alternative choices for which feedbacks are included, but the choice needs to be precisely defined. Otherwise values inferred for Earth system sensitivity may be ambiguous and yield a greater range than dictated by the physics.

We suggest that it is useful to define additional climate sensitivities that build on the fast feedback sensitivity, $S_{ff}$, via sequential addition of slow feedback processes. We focus first on climate sensitivity combining fast feedbacks and slow surface change, $S_{ff+sur}$.

$S_{ff+sur}$ can be evaluated empirically from documented climate changes. Sensitivity $S_{ff+sur}$ is useful for cases in which atmospheric GHG changes are known. We note two specific cases.

One case in which $S_{ff+sur}$ is useful is the era of human-made climate change. Past GHG amounts are known from ice core data and in situ measurements, and future GHG changes can be estimated from GHG emission scenarios and carbon-cycle calculations. A portion of the GHG change is due to slow climate feedbacks, but by specifying observed GHG amounts the GHG effect is included precisely. This approach improves the prospect of assessing other contributions to climate sensitivity, including the surface climate feedback.

A second case in which $S_{ff+sur}$ is useful is $CO_2$ change over millions of years due to plate tectonics. Such long-term $CO_2$ changes, which can be estimated from proxy $CO_2$ measures (Beerling and Royer, 2011) or carbon cycle models (Berner, 2004), are a climate forcing, an imposed perturbation of the planet's energy balance.

Specifically, let us consider $CO_2$ changes during the Cenozoic Era. Earth was so warm in the early Cenozoic (Fig. 1) that there were no large ice sheets. But long-term cooling began about 50 Mya (million years ago), and by about 34 Mya a large ice sheet formed on Antarctica. After further global cooling ice sheets formed in the Northern Hemisphere during the past several million years. An increasing amplitude of temperature oscillations accompanied increasingly large ice sheets in the Pliocene and Pleistocene (Fig. 1b).

Ice sheet changes in the Cenozoic make it clear that climate sensitivity including slow feedbacks is a strong function of the climate state. The growing amplitude of glacial-interglacial oscillations in the Plio-Pleistocene is due to an increasing surface albedo feedback. But surface albedo feedback vanishes as the ice sheets disappear. It follows that climate sensitivity $S_{ff+sur}$ is a function of climate state and the sign (positive or negative) of the climate forcing.

$S_{ff+sur}$ is ~ 1.5°C per W/m$^2$ (6°C for doubled $CO_2$) during the Pleistocene (Hansen et al., 2008). That conclusion is obvious from Fig. 3, which shows that the GHG and surface albedo, as boundary forcings, contribute equally to global temperature change. With both of them considered as boundary forcings, the fast feedback sensitivity is 3°C for doubled $CO_2$. But with GHGs considered to be a forcing, the sensitivity becomes 6°C for doubled $CO_2$.

Sensitivity $S_{ff+sur}$ ~ 1.5°C per W/m$^2$ does not necessarily apply to positive forcings today, because present climate is near the warm extreme of the Pleistocene range. The decreasing amplitude of glacial-interglacial temperature oscillations between the late Pleistocene and Pliocene (Fig. 1b) suggests a substantially smaller $S_{ff+sur}$ for the Holocene-Pliocene climate change than for the Holocene-LGM climate change. Hansen et al. (2008) show that the mean $S_{ff+sur}$ for the entire range from the Holocene to a climate just warm enough to lose the Antarctic ice sheet is almost 1.5°C per W/m$^2$. But most of the surface albedo feedback in that range of climate is associated with loss of the Antarctic ice sheet. Thus the estimate of Lunt et al. (2010), that $S_{ff}$ is increased by a factor of 1.3-1.5 by slow surface feedbacks (reduced ice and increased vegetation cover) for the climate range from the Holocene to the middle Pliocene is consistent with the Hansen et al. (2008) estimate for the mean $S_{ff+sur}$ between 34 Mya and today.

Another definition of Earth system sensitivity with merit is the sensitivity to $CO_2$ change, with accompanying natural changes of non-$CO_2$ GHG changes counted as feedbacks. We could call this the ff+sur+ghg sensitivity (ghg = GHG – $CO_2$), but for brevity we suggest $S_{CO2}$. This sensitivity has the merit that $CO_2$ is the principal GHG forcing and perhaps the only one with good prospects for quantification of its long-term changes. It is likely that non-$CO_2$ trace gases increase as global temperature increases, as found in chemical modeling studies (Beerling et al., 2009, 2011). Non-$CO_2$ GHGs contributed 0.75 W/m$^2$ of the LGM-Holocene forcing, thus amplifying $CO_2$ forcing (2.25 W/m$^2$) by one-third (section S1 of Hansen et al., 2008). GHG and surface boundary forcings co-varied 1-to-1 in the late Pleistocene as a function of temperature (Fig. 5). Thus if non-$CO_2$ trace gases are counted as a fast feedback, the fast-feedback sensitivity becomes 4°C for doubled $CO_2$ and $S_{CO2}$ becomes 1°C per W/m$^2$, for the planet without ice sheets (no slow surface albedo feedback). $S_{CO2}$ from the Holocene as initial state is thus 8°C for doubled $CO_2$ and 2°C per W/m$^2$ for negative forcings; $S_{CO2}$ is samller for a positive forcing, but it is nearly that large for a positive forcing just large enough to melt the Antarctic ice sheet. $S_{CO2}$ is the definition of Earth system sensitivity used by Royer et al. (2011), which substantially accounts for the high sensitivities that they estimate.

When climate sensitivity is inferred empirically from long-term climate change and GHG changes, it is necessary to include the effect of other changing boundary forcings, such as solar irradiance and continental locations, if the changes are substantial.  However, such changes are negligible for a rapid change of GHGs as in the Paleocene-Eocene Thermal Maximum.

Table 1.  Climate sensitivities, which are equilibrium responses to a specified forcing.

| Name, Explanation | Estimated Value | Comments |
|---|---|---|
| $S_{ff}$, all fast feedbacks including aerosols | 0.75°C per W/m$^2$<br>3°C for 2×$CO_2$ | valid for positive and negative forcings from current climate |
| $S_{ff+sur}$, fast feedbacks plus surface feedbacks | 1.5°C per W/m$^2$<br>6°C for 2×$CO_2$ | valid for negative forcing from Holocene climate state; value is less for positive forcing (see text) |
| $S_{CO2}$, specified $CO_2$ amount as forcing | 2°C per W/m$^2$<br>8°C for 2×$CO_2$ | valid for negative forcing from Holocene climate state; value is less for positive forcing (see text) |
| $S_{ff+sf}$, fast feedbacks plus surface and GHG feedbacks | remarkably large, especially for negative forcings | for $CO_2$ forcing, the long climate response time for high sensitivity implies that negative (diminishing) feedbacks will be important |

    The ultimate Earth system sensitivity is $S_{ff+sf}$, the sensitivity including all fast and slow feedbacks, i.e., surface feedbacks and all GHG feedbacks including $CO_2$.  $S_{ff+sf}$ is relevant to changing solar irradiance, for example.  Apparently $S_{ff+sf}$ is remarkably large in the late Pleistocene.  However, the extreme sensitivity implied by late Pleistocene climate oscillations was associated with a cooling climate that caused the surface (ice sheet) albedo feedback to be the largest it has been since perhaps the early Permian, about 300 million years ago (Royer, 2006).  Given human-made GHGs, including movement of fossil carbon into surface reservoirs, the extreme $S_{ff+sf}$ of the late Pleistocene will not be relevant as long as humans exist.

    In principle $S_{ff+sf}$ is relevant for interpretation of past climate change due to Earth orbital forcing.  However, Earth orbital forcing is subtle and complex.  Useful applications will require definition of an appropriate effective forcing, i.e., a forcing that incorporates the efficacy (Hansen et al., 2005) of the orbital forcing as a function of latitude and season.

    In conclusion, which sensitivity, if any, deserves the moniker 'Earth system sensitivity'?  From an academic perspective, $S_{ff+sf}$ is probably the best choice.   From a practical perspective $S_{ff}$ and $S_{ff+sur}$ are both needed for analysis of human-made climate change.  From a paleoclimate perspective, $S_{CO2}$ is very useful.  So there is more than one useful choice.  The important point is to make clear exactly what is meant.  And remember to specify the reference climate state.  Table 1 summarizes alternative climate sensitivities.

## 4. What is the dangerous level of global warming?

Paleoclimate data yield remarkably rich and precise information on climate sensitivity. We suggest that paleoclimate data on climate change and climate sensitivity can be pushed further to yield an accurate evaluation of the dangerous level of global warming.

Broad-based assessments, represented by a "burning embers" diagram in IPCC (2001, 2007), suggested that major problems begin with global warming of 2-3°C relative to global temperature in year 2000. Sophisticated probabilistic analyses (Schneider and Mastrandrea, 2005) found a median "dangerous" threshold of 2.85°C above global temperature in 2000, with the 90 percent confidence range being 1.45-4.65°C.

The IPCC analyses contributed to a European Union (2008) decision to support policies aimed at keeping global warming less than 2°C relative to pre-industrial times (1.3°C relative to the 11-year running mean global temperature in 2000). Subsequent documents of the European Union (2010) and a group of Nobel laureates (Stockholm Memo, 2011) reaffirm this 2°C target.

We will suggest, however, that paleoclimate data imply that 2°C global warming would be a disaster scenario for much of humanity and many other species on the planet.

Prior interglacial periods that were warmer than the Holocene can play a key role in assessing the dangerous level of global warming. As shown in Fig. 2d,e, the interglacials peaking near 125 and 400 ky ago (Eemian and Holsteinian, known in paleoclimate literature as Marine Isotope Stages 5e and 11, respectively) were warmer than the Holocene. However, the ice cores and ocean cores do not seem to agree on how warm those prior interglacials were. So we must first consider the differences between these two paleoclimate records.

### 4.1 Ice cores versus ocean cores

The Antarctic Dome C ice core, with the approximation that global temperature change on millennial time scales is half as large as polar temperature change, indicates that the Eemian and Holsteinian may have been 1 to 2°C warmer than the Holocene (Fig. 2d). However, the ocean core record (Fig. 2e) suggests that these interglacial periods were only a fraction of a degree warmer than the Holocene. Assessment of dangerous global warming requires that we understand the main reasons for these different pictures, and achieving that objective requires discussion of the nature of these two different records.

*Ice cores.* $H_2O$ isotope amounts in the polar ice cores depend upon the air temperature where and when the snowflakes formed above the ice sheets. Several adjustments[3] to the ice core temperature record have been suggested with the aim of producing a more homogeneous record, i.e., a result that more precisely defines the surface air temperature change at a fixed location and fixed altitude. However, these adjustments are too small to remove the discrepancy that exists when global temperature inferred from ice cores is compared with either ocean core temperature change (Fig. 2e) or with our calculations based on greenhouse gas and albedo climate forcings (Fig. 2d).

---

[3] One adjustment accounts for estimated glacial-interglacial change of the source region for the water vapor that forms the snowflakes (Vimeux et al., 2002). The source location depends on sea ice extent. This correction reduces interglacial warmth and thus reduces the discrepancy with the calculated interglacial temperatures in Fig. 4a.

Another adjustment accounts for change of ice sheet thickness (Masson-Delmotte et al., 2010). This adjustment increases the fixed-altitude temperature in the warmest interglacials. The correction is based on ice sheet models, which yield a greater altitude for the central part of the ice sheet, even though sea level was higher in these interglacials and thus ice sheet volume was smaller. This counter-intuitive result is conceivable because snowfall is greater during warmer interglacials, which could make the central altitude greater despite the smaller ice sheet volume. But note that the correction is based on ice sheet models that may be "stiffer" than real-world ice sheets.

The principal issue about temperature change on top of the ice sheet during the warmest interglacials is whether the simple (factor of two) relationship with global mean temperature change is accurate during the warmest interglacials. That simple prescription works well for the Holocene and for all the glacial-interglacial cycles during the early part of the 800,000 year record, when the interglacials were no warmer than the Holocene.

We suggest that interglacial periods warmer than the Holocene, such as the Eemian, had moved into a regime in which there was less summer sea ice around Antarctica and Greenland, there was summer melting on the lowest elevations of the ice sheets, and there was summer melting on the ice shelves, which thus largely disappeared. In such a regime, even small global warming above the level of the Holocene could generate disproportionate warming on the Antarctic and Greenland ice sheets, more than double the global mean warming.

Summer melting on lower reaches of the ice sheets and on ice shelves introduces the "albedo flip" mechanism (Hansen et al., 2007a). This phase change of water causes a powerful local feedback, which, together with moderate global warming, can increase the length of the melt season. Increased warm season melting increases the ice sheet temperature and affects sea level on a time scale that is being debated, as discussed below. Increased surface melting, loss of ice shelves, and reduced summer sea ice around Antarctica and Greenland would have a year-round effect on temperature over the ice sheets. Indeed, more open water increases heat flow from ocean to atmosphere with the largest impact on surface air temperature in the cool seasons.

We interpret the stability of Holocene sea level as a consequence of the fact that global temperature was just below the level required to initiate the "albedo flip" mechanism on the fringes of West Antarctica and on most of Greenland. An important implication of this interpretation is that the world today is on the verge of, or has already reached, a level of global warming for which the equilibrium surface air temperature response on the ice sheets will exceed global warming by much more than a factor of two. Below we cite empirical evidence in support of this interpretation. First, however, we must discuss limitations of ocean core data.

*Ocean cores.* Extraction of surface temperature from ocean cores has its own problems. Although obtained from many sites, the deep ocean data depend mainly on surface temperature at high latitude regions of deep water formation that may move as climate changes. As climate becomes colder, for example, sea ice expands and the location of deep water formation may move equatorward. Fortunately, the climates of most interest range from the Holocene toward warmer climates. Because of geographical constraints it seems unlikely that the present sites of deep water formation would move much in response to moderate global warming.

A second problem with ocean cores is that deep ocean temperature change is limited as ocean water nears its freezing point. That is why deep ocean temperature change between the LGM and the Holocene was only two-thirds as large as global average surface temperature change. However, by using a constant adjustment factor (1.5) in Fig. 2, based on the LGM to Holocene climate change, we overstate this magnification at interglacial temperatures and understate the magnification at the coldest climates, thus maximizing the possibility for the deep ocean temperature to reveal (and exaggerate) interglacial warmth. Yet no interglacial warm spikes appear in the ocean core temperature record (Fig. 2e).

A third issue concerns the temporal resolution of ocean cores. Bioturbation, i.e., mixing of ocean sediments by worms, smoothes the ocean core record, especially at locations where ocean sediments accumulate slowly. However, the interglacial periods of primary concern, the Eemian and Holsteinian, were longer than the resolution limit of most ocean cores.

We conclude that ocean cores provide a better measure of global temperature change than ice cores during those interglacial periods that were warmer than the pre-industrial Holocene.

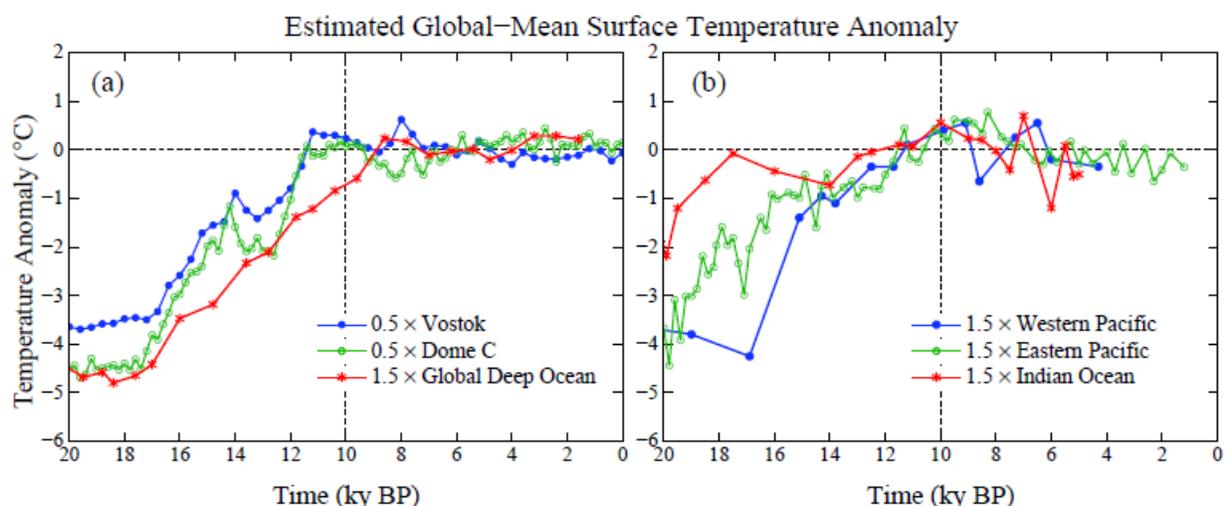

**Fig. 5.** Estimates of global temperature change inferred from Antarctic ice cores (Vimeux et al., 2002; Jouzel et al., 2007) and ocean sediment cores (Medina-Elizade and Lea, 2005; Lea et al., 2000, 2006; Saraswat et al., 2005). Zero-point temperature is the mean for the past 10 ky.

### 4.2 Holocene versus prior interglacial periods and the Pliocene

How warm is the world today relative to peak Holocene temperature? Peak Holocene warmth, is commonly placed about 8,000 years ago, but it varies from one place to another (Mayewski et al., 2004). Our interest is global mean temperature, not regional variations.

Fig. 5 compares several temperature records for the sake of examining Holocene temperature change. Zero temperature is defined as the mean for the past 10,000 years. The records are made to approximate global temperature by dividing polar temperatures by two and multiplying deep ocean and tropical ocean mixed layer[4] temperature by a factor 1.5. Fig. 5 indicates that global temperature has been relatively stable during the Holocene.

So how warm is it today relative to peak Holocene warmth? Fig. 5, especially the global deep ocean temperature, shows that the world did not cool much in the Holocene. Consistent with our earlier study (Hansen et al., 2006), we conclude that, with the global surface warming of 0.7°C between 1880 and 2000 (Hansen et al., 2010), global temperature in year 2000 has reached at least the Holocene maximum.

How does peak Holocene temperature compare with prior warmer interglacial periods, specifically the Eemian and Holsteinian interglacial periods, and with the Pliocene?

Fig. 6 shifts the temperature scale so that it is zero at peak Holocene warmth. The temperature curve is based on the ocean core record of Fig. 1 but scaled by the factor 1.5, which is the scale factor relevant to the total LGM-Holocene climate change. Thus for climates warmer than the Holocene, Fig. 6 may <u>exaggerate</u> actual temperature change.

One conclusion deserving emphasis is that global mean temperatures in the Eemian and Holsteinian were less than 1°C warmer than peak Holocene global temperature. Therefore, these interglacial periods were also less than 1°C warmer than global temperature in year 2000.

Fig. 6 also suggests that global temperature in the early Pliocene, when sea level was about 25 m higher than today (Dowsett et al., 1994), was only about 1°C warmer than peak Holocene temperature, thus 1-2°C warmer than recent (pre-industrial) Holocene. That conclusion requires a caveat about possible change of location of deepwater formation, stronger than the same caveat in comparing recent interglacial periods. Substantial change in the location of deep water formation is more plausible in the Pliocene because of larger Arctic warming at that time (Dowsett et al., 1999); also ocean circulation may have been altered in the early Pliocene by closure of the Panama Seaway, although the timing of that closure is controversial (Haug and Tiedemann, 1998).

---

[4] Indian and Pacific Ocean temperatures in Fig. 5 are derived from forams that lived in the upper ocean, as opposed to benthic forams used to obtain global deep ocean temperature. The Eastern Pacific temperature in Fig. 5b is the average for two locations, north and south of the equator, which are shown individually by Hansen et al. (2006).

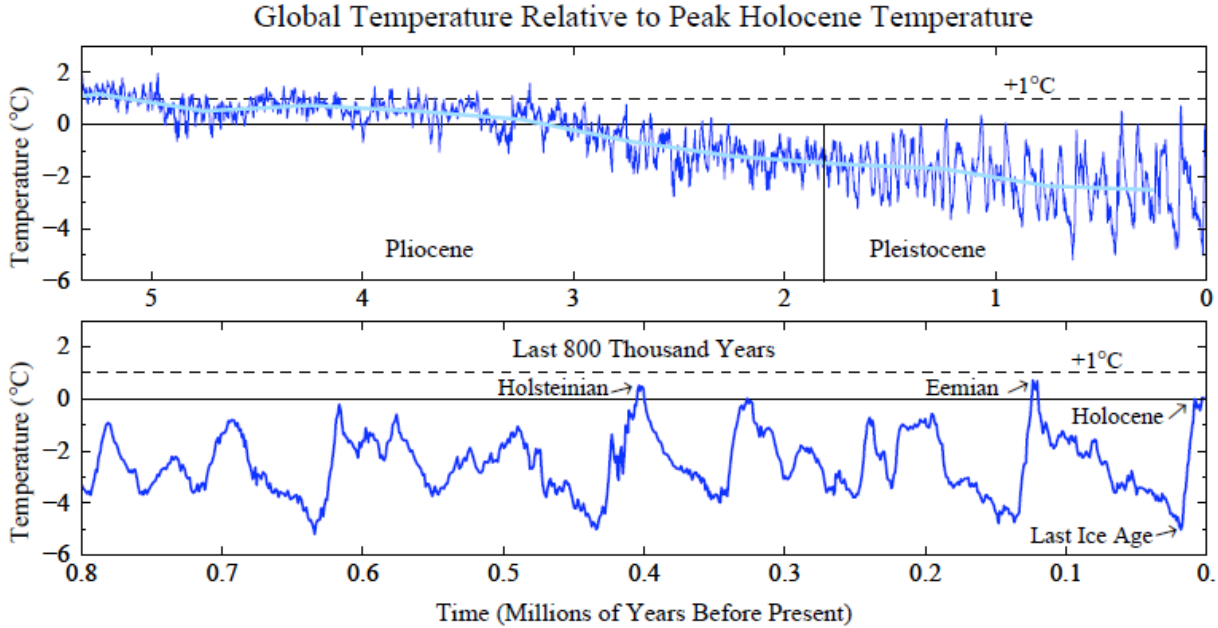

**Fig. 6.** Global temperature relative to peak Holocene temperature, based on ocean core records in Fig. 1. Deep ocean temperature change is amplified by factor 1.5 to obtain this estimate of surface change.

Is such small Pliocene warming inconsistent with PRISM (Pliocene Research, Interpretation and Synoptic Mapping Project) reconstructions of mid-Pliocene (3-3.3 My ago) climate (Dowsett et al., 1996, 2009 and references therein)? Global mean surface temperatures in climate models forced by PRISM boundary conditions yield global warming of about 3°C (Lunt et al., 2010) relative to pre-industrial climate. However, it must be borne in mind that "PRISM's goal is a reconstruction of a 'super interglacial', not mean conditions" (Dowsett et al., 2009), which led to (intentional, as documented) choices of the warmest conditions in a variety of data sets that were not necessarily well correlated in time.

Perhaps the most striking characteristic of Pliocene climate reconstructions is that low latitude ocean temperatures were similar to those today, except that the east-west temperature gradient was reduced in the tropical Pacific Ocean, possibly resembling permanent El Nino conditions (Wara et al., 2005). High latitudes were warmer than today, the ice sheets smaller, and sea level about 25 m higher (Dowsett et al., 2009; Rohling et al., 2009). Atmospheric $CO_2$ amount was larger in the Pliocene, recent estimates being 390 ± 25 ppm (Pagani et al., 2009) and 365 ± 35 ppm (Seki et al., 2010). It is likely that both elevated $CO_2$ and increased poleward heat transports by the ocean and atmosphere contributed to large high latitude warming, but Pliocene climate has not been well simulated from first principles by climate models. Indeed, today's climate models generally are less sensitive to forcings than the real world (Valdes, 2011), suggesting that models do not capture well some amplifying climate feedbacks and thus making empirical assessment via Earth's history of paramount importance.

We conclude that Pliocene temperatures probably were no more than 1-2°C higher on global average than peak Holocene temperature. Regardless of precise Pliocene temperatures, the extreme polar warmth and diminished ice sheets in the Pliocene are consistent with the picture we painted above: Earth today, with global temperature having returned to at least the Holocene maximum, is poised to experience strong amplifying polar feedbacks in response to even modest additional global mean warming.

### 4.3 Sea level

Sea level rise potentially sets a low limit on the dangerous level of global warming. Civilization developed during a time of unusual climate stability and sea level stability. Much of the world's population and infrastructure are located along coastlines.

Sea level rise, despite its potential importance, is one of the least well understood impacts of human-made climate change. The difficulty stems from the fact that ice sheet disintegration is a complex non-linear phenomenon that is inherently difficult to simulate, as well as from the absence of a good paleoclimate analogue for the rapidly increasing human-made climate forcing. Here we try to glean information from several different sources.

*Paleoclimate data.* Fig. 4 shows that the equilibrium (eventual) sea level change in response to global temperature change is about 20 meters for each degree Celsius global warming. (The variable in Fig. 4 is the albedo forcing due to change of ice sheet size, but albedo forcing and sea level change are proportional; cf. Fig. S4 of Hansen et al., 2008).

This relationship, an equilibrium sea level rise of 20 meters per degree Celsius, continues to be valid for warmer climates. Fig. 6 shows that average temperature in the early Pliocene, when sea level was of the order of 20 m higher than today, was about 1°C above peak Holocene temperature. Fig. 1 shows that just prior to Antarctic glaciation, 34 million years ago, global temperature was at most about 3°C above peak Holocene temperature and sea level must have been at least 60 meters higher because there were no large ice sheets on the planet.

We conclude that eventual sea level rise of several tens of meters must be anticipated in response to the global warming of several degrees Celsius that is expected under business-as-usual (BAU) climate scenarios (IPCC, 2001, 2007).

Paleoclimate data are less helpful for estimating the expected <u>rate</u> of sea level rise. Besides the lack of a good paleo analog to the rapid human-made forcing, the dating of paleoclimate changes is imprecise. Hansen et al. (2007a) conclude that there is no discernable lag between climate forcing (Northern Hemisphere late spring insolation maximum) and the maximum rate of sea level rise for the two deglaciations that are most accurately dated. Thus they argue that it does not require millennia for substantial ice sheet response to a forcing, but the weak, slowly changing paleoclimate forcing prevents a more quantitative conclusion.

*Sea level change estimates for 21$^{st}$ century.* IPCC (2007) projected sea level rise by the end of this century of about 29 cm (midrange 20-43 cm, full range 18-59 cm). These projections did not include contributions from ice sheet dynamics, on the grounds that ice sheet physics is not understood well enough.

Rahmstorf (2007) made an important contribution to the sea level discussion by pointing out that even a linear relation between global temperature and the rate of sea level rise, calibrated with 20$^{th}$ century data, implies a 21$^{st}$ sea level rise of about a meter, given expected global warming for BAU greenhouse gas emissions. Vermeer and Rahmstorf (2009) extended Rahmstorf's semi-empirical approach by adding a rapid response term, projecting sea level rise by 2100 of 0.75-1.9 m for the full range of IPCC climate scenarios. Grinsted et al. (2010) fit a 4-parameter linear response equation to temperature and sea level data for the past 2000 years,

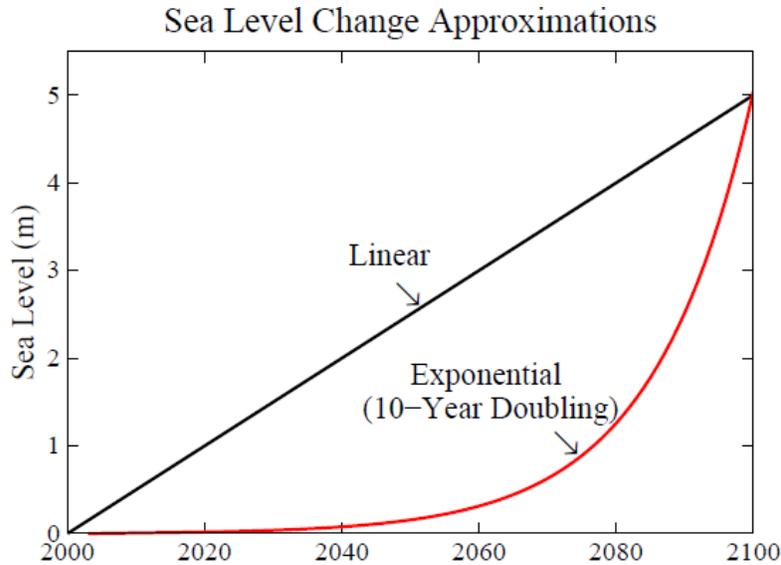
**Fig. 7.** Five-meter sea level change in 21$^{st}$ century under assumption of linear change and exponential change (Hansen, 2007), the latter with a 10-year doubling time.

projecting a sea level rise of 0.9-1.3 m by 2100 for a middle IPCC scenario (A1B).  These projections are typically a factor of 3-4 larger than the IPCC (2007) estimates, and thus they altered perceptions about the potential magnitude of human-caused sea level change.

Alley (2010) reviewed projections of sea level rise by 2100, showing several clustered around 1 m and one outlier at 5 m, all of these approximated as linear in his graph.  The 5 m estimate is what Hansen (2007) suggested was possible under IPCC's BAU climate forcing.  Such a graph is comforting – not only does the 5-meter sea level rise disagree with all other projections, but its half-meter sea level rise this decade is clearly preposterous.

However, the fundamental issue is linearity versus non-linearity.  Hansen (2005, 2007) argues that amplifying feedbacks make ice sheet disintegration necessarily highly non-linear, and that IPCC's BAU forcing is so huge that it is difficult to see how ice shelves would survive.  As warming increases, the number of ice streams contributing to mass loss will increase, contributing to a nonlinear response that should be approximated better by an exponential than by a linear fit.  Hansen (2007) suggested that a 10-year doubling time was plausible, and pointed out that such a doubling time, from a 1 mm per year ice sheet contribution to sea level in the decade 2005-2015, would lead to a cumulative 5 m sea level rise by 2095.

Nonlinear ice sheet disintegration can be slowed by negative feedbacks.  Pfeffer et al. (2008) argue that kinematic constraints make sea level rise of more than 2 m this century physically untenable, and they contend that such a magnitude could occur only if all variables quickly accelerate to extremely high limits.  They conclude that more plausible but still accelerated conditions could lead to sea level rise of 80 cm by 2100.

The kinematic constraint may have relevance to the Greenland ice sheet, although the assumptions of Pfeffer at al. (2008) are questionable even for Greenland.  They assume that ice streams this century will disgorge ice no faster than the fastest rate observed in recent decades.  That assumption is dubious, given the huge climate change that will occur under BAU scenarios, which have a positive (warming) climate forcing that is increasing at a rate dwarfing any known natural forcing.  BAU scenarios lead to $CO_2$ levels higher than any since 32 My ago, when Antarctica glaciated.  By mid-century most of Greenland would be experiencing summer melting in a longer melt season. Also some Greenland ice stream outlets are in valleys with bedrock below sea level. As the terminus of an ice stream retreats inland, glacier sidewalls can collapse, creating a wider pathway for disgorging ice.

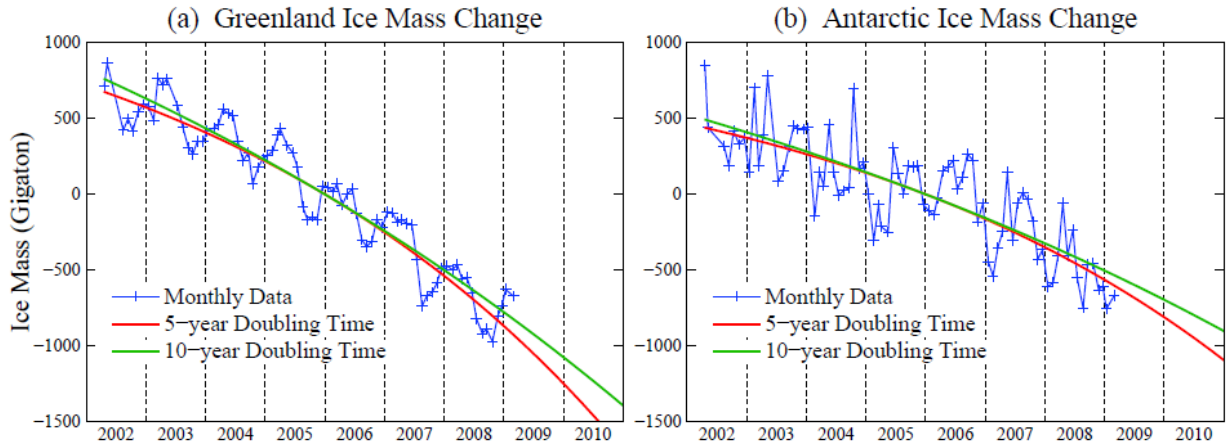

**Fig. 8.** Greenland (a) and Antarctic (b) mass change deduced from gravitational field measurements by Velicogna (2009) and best-fits with 5-year and 10-year mass loss doubling times.

The main flaw with the kinematic constraint concept is the geology of Antarctica, where large portions of the ice sheet are buttressed by ice shelves that are unlikely to survive BAU climate scenarios. West Antarctica's Pine Island Glacier (PIG) illustrates nonlinear processes already coming into play. The floating ice shelf at PIG's terminus has been thinning in the past two decades as the ocean around Antarctica warms (Shepherd et al., 2004; Jenkins et al., 2010). Thus the grounding line of the glacier has moved inland by 30 km into deeper water, allowing potentially unstable ice sheet retreat. PIG's rate of mass loss has accelerated almost continuously for the past decade (Wingham et al., 2009) and may account for about half of the mass loss of the West Antarctic ice sheet, which is of the order of 100 km$^3$ per year (Sasgen et al., 2010).

PIG and neighboring glaciers in the Amundsen Sea sector of West Antarctica, which are also accelerating, contain enough ice to contribute 1-2 m to sea level. Most of the West Antarctic ice sheet, with at least 5 m of sea level, and about a third of the East Antarctic ice sheet, with another 15-20 m of sea level, are grounded below sea level. This more vulnerable ice may have been the source of the 25 ± 10 m sea level rise of the Pliocene (Dowsett et al., 1990, 1994). If human-made global warming reaches Pliocene levels this century, as expected under BAU scenarios, these greater volumes of ice will surely begin to contribute to sea level change. Indeed, satellite gravity and radar interferometry data reveal that the Totten Glacier of East Antarctica, which fronts a large ice mass grounded below sea level, is already beginning to lose mass (Rignot et al., 2008).

The eventual sea level rise due to expected global warming under BAU GHG scenarios is several tens of meters, as discussed at the beginning of this section. From the present discussion it seems that there is sufficient readily available ice to cause multi-meter sea level rise this century, if dynamic discharge of ice increases exponentially. Thus current observations of ice sheet mass loss are of special interest.

*Ice sheet mass loss.* The best indication and quantification of possible non-linear behavior will be precise measurements of ice sheet mass change. Mass loss by the Greenland and Antarctic ice sheets can be deduced from satellite measurements of Earth's gravity field. Fig. 8 shows the changing mass of both ice sheets as reported by Velicogna (2009).

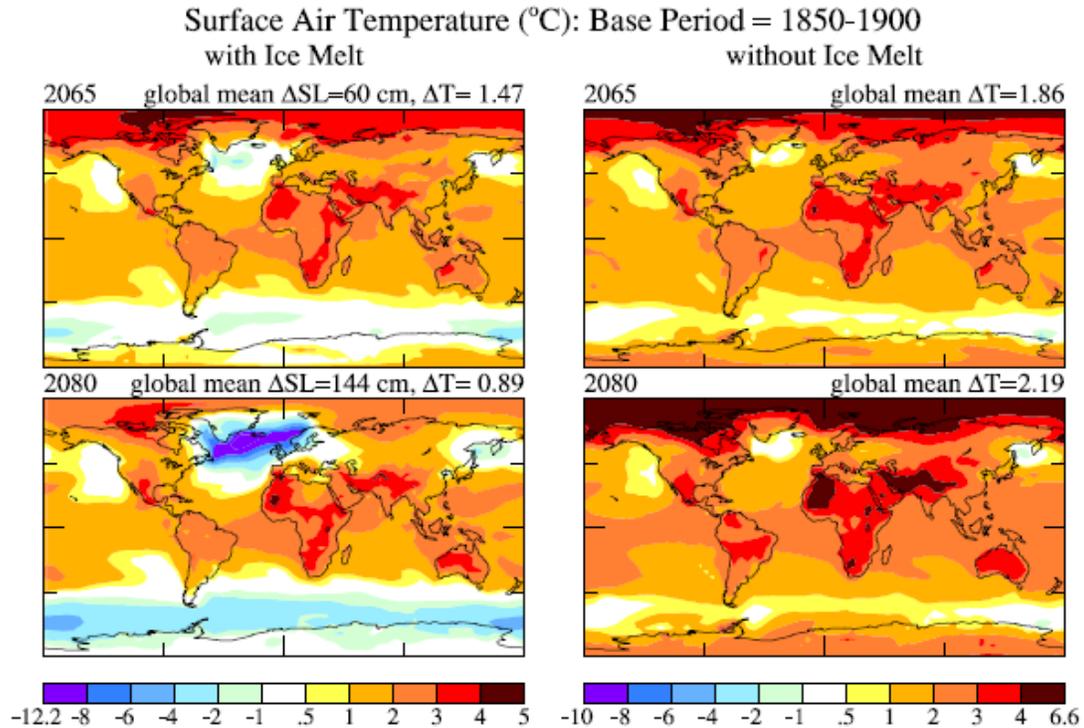

**Fig. 9.** Surface air temperature change in 2065 (above) and 2080 (below) relative to 18501900 in simulations with GISS climate model using IPCC A1B scenario. Maps on left include ice melt, which is put half into the North Atlantic and half into the Southern Ocean, with ice melt doubling every ten years.

These data records suggest that the rate of mass loss is increasing, indeed nearly doubling over the period of record, but the record is too short to provide a meaningful evaluation of a doubling time. Also there is substantial variation among alternative analyses of the gravity field data (Sorensen and Forsberg, 2010), although all analyses have the rate of mass loss increasing over the period of record.

We conclude that available data for the ice sheet mass change are consistent with our expectation of a non-linear response, but the data record is too short and uncertain to allow quantitative assessment. A 10-year doubling time, or even shorter, is consistent with the gravity field data, but because of the brevity of the record even a linear mass loss cannot be ruled out. Assessments will rapidly become more meaningful in the future, if high-precision gravity measurements are continued.

*Iceberg cooling effect.* Exponential change cannot continue indefinitely. The negative feedback terminating exponential growth of ice loss is probably regional cooling due to the thermal and fresh-water effects of melting icebergs. Temporary cooling occurs as icebergs and cold fresh glacial melt-water are added to the Southern Ocean and the North Atlantic Ocean.

As a concrete example, Fig. 9 shows the global temperature change in simulations with GISS modelE (Schmidt et al., 2006; Hansen et al., 2007c) with and without the melting iceberg effect. GHGs follow the A1B scenario, an intermediate business-as-usual scenario (IPCC, 2001, 2007; see also Figs. 2 and 3 of Hansen et al., 2007b). Ice melt rate is such that it contributes 1mm/year to sea level in 2010, increasing with a 10-year doubling time; this melt rate constitutes 0.034 Sv (1 Sverdrup = 1 million m$^3$ per second) in 2065 and 0.1 Sv in 2080. Half of this meltwater is added in the North Atlantic and half in the Southern Ocean.

By 2065, when the sea level rise (from ice melt) is 60 cm relative to 2010, the cold freshwater reduces global mean warming (relative to 1880) from 1.86°C to 1.47°C. By 2080, when sea level rise is 1.4 m, global warming is reduced from 2.19°C to 0.89°C. These experiments are described in a paper in preparation (Hansen, Ruedy and Sato, 2011), which includes other GHG scenarios, cases with ice melt in one hemisphere but not the other, and investigation of the individual effects of freshening and cooling by icebergs (the freshening is more responsible for the reduction of global warming). Note that the magnitude of the regional cooling is comparable to that in 'Heinrich' events in the paleoclimate record (Bond et al., 1992), these events involving massive iceberg discharge at a rate comparable to that in our simulations. Given that the possibility of sea level rise of the order of a meter is now widely accepted, it is important that simulations of climate for the 21$^{st}$ century and beyond include the iceberg cooling effect.

Detailed consideration of the climate effects of freshwater from ice sheet disintegration, which has a rich history (Broecker et al., 1990; Rahmstorf, 1996; Manabe and Stouffer, 1997), is beyond the scope of our present paper. However, we note that the temporary reduction of global warming provided by icebergs is not likely to be a blessing. Stronger storms driven by increased latitudinal temperature gradients, combined with sea level rise, likely will produce global havoc. It was the prospect of increased ferocity of continental-scale frontal storms, with hurricane-strength winds powered by the contrast between air masses cooled by ice melt and tropical air that is warmer and moister than today, which gave rise to the book title "Storms of My Grandchildren" (Hansen, 2009).

## 5. Discussion

Earth's paleoclimate history is remarkably rich in information on how sensitive climate is to forcings, both natural forcings and human-made forcings. Huge glacial-to-interglacial climate swings have been driven by very weak climate forcings, as the climate response is amplified by both fast feedbacks, such as water vapor and aerosols, and slow feedbacks, especially $CO_2$ and surface albedo. The paleoclimate record allows us to deduce that the fast-feedback climate sensitivity is about 3°C global warming for doubled $CO_2$. Climate sensitivity including slow feedbacks depends upon the initial climate state, but it is generally much greater than the fast-feedback climate sensitivity.

Carbon dioxide functions as an amplifying slow climate feedback, because the division of $CO_2$ among its surface reservoirs (atmosphere, ocean, soil, biosphere) shifts toward more $CO_2$ in the atmosphere as the planet becomes warmer. However, $CO_2$ is also a climate forcing when it is extracted from the solid earth and injected into the surface reservoirs either by enhanced volcanic activity or by humans burning fossil fuels. The $CO_2$ so extracted from the deep Earth remains in the surface reservoirs for millennia, until the weathering process eventually results in deposition of carbonates on the ocean floor. Thus the slow $CO_2$ and albedo feedbacks, as well as the fast feedbacks, will eventually have time to respond to human-made fossil fuel $CO_2$ emissions.

The paleoclimate record is also a good source of information on the level of global warming that will eventually yield a markedly different planet than the one on which civilization developed. Paleoclimate data help us assess climate sensitivity and potential human-made climate effects. We conclude that Earth in the warmest interglacial periods of the past million years was less than 1°C warmer than in the Holocene. Polar warmth in those interglacials and in the Pliocene does not imply that a substantial cushion remains between today's climate and dangerous warming, but rather that Earth is poised to experience strong amplifying polar feedbacks in response to moderate additional global warming.

## 5.1. How warm were recent interglacial periods and the Pliocene?

There are numerous statements and presumptions in the scientific literature that prior interglacial periods such as the Eemian were as much as a few degrees warmer than the Holocene (e.g., Rohling et al., 2008; Church et al., 2010), and this perception has probably influenced estimates of what constitutes a dangerous level of global warming. These perceptions about interglacial global temperature must derive at least in part from the fact that Greenland and Antarctica did achieve such higher temperatures during the Eemian.

However, we interpret these temperatures on the ice sheets as being local and unrepresentative of global mean temperature anomalies. The polar ice sheet temperature anomalies were likely magnified by the fact that these warmer interglacial periods had little summer sea ice or ice shelves around the Greenland and Antarctic continents.

We argue that global deep ocean temperatures provide a better measure of global mean temperature anomalies than polar ice cores during the interglacial periods. Ocean cores have a systematic difficulty as a measure of temperature change when the deep ocean temperature approaches the freezing point, as quantified by Waelbroeck et al. (2002). However, in using the known surface temperature change between the last glacial maximum and the Holocene as an empirical calibration, we maximize (i.e., we tend to exaggerate) the ocean core estimate of global surface warming during warmer interglacials relative to the Holocene.

Ocean core data is also affected by the location of deep water formation, which may change. However, the location of deep water formation around Antarctica, which affects deep Pacific Ocean temperature, is limited by the Antarctic geography and is unlikely to be shifted substantially in interglacial periods warmer than the Holocene.

Fig. 2 provides unambiguous discrimination between ice and ocean core measures of global temperature change. Climate forcings for the past 800,000 years are known accurately. Climate sensitivity cannot vary much from one interglacial period to another. Ocean core temperatures give a consistent climate sensitivity for the entire 800,000 years. In contrast, ice core temperature (Fig. 2d) leads to the illogical result that climate sensitivity depends on time.

We conclude that ocean core data are correct in indicating that global surface temperature was only slightly higher in the Eemian and Holsteinian interglacial periods than in the Holocene, at most by about 1°C, but probably by only several tenths of a degree Celsius. By extension (see Fig. 6), the Pliocene was at most 1-2°C warmer than the Holocene on global mean.

## 5.2. How slow are slow feedbacks?

Observed time scales of GHG and surface albedo variability (Fig. 2) are the time scales of orbital variations, thus not necessarily an internal time scale of the feedback processes. Indeed, we do not expect slow feedbacks to be inherently <u>that</u> slow. We have argued (Hansen, 2005; Hansen et al., 2007a) that the ice sheet response to a strong rapid forcing is much faster than the time scale of orbital changes, with substantial response likely within a century.

Debating what sea level will be on a specific date such as 2100, however, misses an important point concerning response times. The carbon cycle response time, i.e., the time required for $CO_2$ from fossil fuel burning to be removed from the surface carbon reservoirs is many millennia (Berner, 2004; Archer, 2005). The ice sheet response time is clearly shorter than this carbon cycle response time, in view of the absence of a discernable lag between paleoclimate forcings and the maximum rate of ice sheet disintegration (Hansen et al., 2007a) and in view of the fact that ice sheet disintegration proceeds at rates up to several meters of sea level rise per century (Fairbanks, 1989) even in response to weak paleoclimate forcings.

Thus burning all or most fossil fuels guarantees tens of meters of sea level rise, as we have shown that the eventual sea level response is about 20 meters of sea level for each degree Celsius of global warming. We suggest that ice sheet disintegration will be a nonlinear process, spurred by an increasing forcing and by amplifying feedbacks, which is better characterized by a doubling time for the rate of mass disintegration, rather than a linear rate of mass change. If the doubling time is as short as a decade, multi-meter sea level rise could occur this century. Observations of mass loss from Greenland and Antarctica are too brief for significant conclusions, but they are not inconsistent with a doubling time of a decade or less. The picture will become clearer as the measurement record lengthens.

There are physical constraints and negative feedbacks that may limit nonlinear ice sheet mass loss. An ice sheet sitting primarily on land above sea level, such as most of Greenland, may be limited by the speed at which it can deliver ice to the ocean via outlet glaciers. But much of the West Antarctic ice sheet, resting on bedrock below sea level, is not so constrained.

We recognize the negative feedback that comes into play as iceberg discharge reaches a rate that cools the regional ocean surface. But that negative feedback would be cold comfort. High latitude cooling and low latitude warming would drive more powerful mid-latitude cyclonic storms, including more frequent cases of hurricane force winds. Such storms, in combination with rising sea level, would be disastrous for many of the world's great cities and they would be devastating for the world's economic well-being and cultural heritage.

## 5.3. How much warming is too much?

The most substantial political effort to place a limit on global warming has been the European Union's target to keep global temperature from exceeding the preindustrial level by more than 2°C (European Union, 2008). This goal was later reaffirmed (European Union, 2010) and it was endorsed by a group of Nobel Laureates in the Stockholm Memo (2011).

However, based on evidence presented in this paper a target of 2°C is not safe or appropriate. Global warming of 2°C would make Earth much warmer than in the Eemian, when sea level was 4-6 meters higher than today. Indeed, with global warming of 2°C Earth would be headed back toward Pliocene-like conditions.

Conceivably a 2°C target is based partly on a perception of what is politically realistic, rather than a statement of pure science. In any event, our science analysis suggests that such a target is not only unwise, but likely a disaster scenario.

Detailed consideration of targets is beyond the scope of this paper, but we note that our present study is consistent with the "target $CO_2$" analysis of Hansen et al. (2008). Those authors argued that atmospheric $CO_2$ should be rolled back from its present ~390 ppm at least to the level of approximately 350 ppm. With other climate forcings held fixed, $CO_2$ at 350 ppm would restore the planet's energy balance and keep human-made global warming less than 1°C, as we and several colleagues discuss in two papers ("Earth's Energy Imbalance" and "The Case for Young People and Nature") in preparation.

**Acknowledgments.** We thank referee (Dana Royer) for helpful suggestions, Gerry Lenfest (Lenfest Foundation), Lee Wasserman (Rockefeller Family Foundation), Stephen Toben (Flora Family Foundation) and NASA program managers Jack Kaye and David Considine for research support, and Gavin Schmidt, Pushker Kharecha, Richard Alley, Christopher Barnet, Peter Barrett, Phil Blackwood, John Breithaupt, Tim Dean, Bruce Edwards, J. Gathright, Michael Le Page, Robert Maginnis, Jon Parker, Tom Parrett, Les Porter, Warwick Rowell, Ken Schatten, Colin Summerhayes and Bart Verheggen for comments on a draft version of this paper.